\documentclass[a4paper,12pt,oneside]{article}
\usepackage[T2A]{fontenc}
\usepackage[cp1251]{inputenc}
\usepackage[english]{babel}
\usepackage{amsmath}
\usepackage{amsmath,mathtools}
\usepackage{amssymb}
\usepackage{graphicx}
\begin{document}

{\Large\bf STEADY AND PERIODIC REGIMES }
\vspace{4pt}

{\Large\bf OF LAMINAR FLOW AROUND THE }

\vspace{4pt}
{\Large\bf ROTATING CYLINDER}

\bigskip
$\mathbf{E.~I.~Kalinin^* ~\& ~A.~B.~Mazo}$
\bigskip

{\it Kazan Federal University, $18$ Kremlyovskaya Street, Kazan $420008$,}

{\it Republic of Tatarstan, Russian Federation}
\bigskip

$^*$Address all correspondence to E.~I.~Kalinin E-mail: kalininEI@yandex.ru

\bigskip

{\it\small\indent
A numerical study of the problem of laminar infinite flow of viscous \hyphenpenalty=10000 incompressible

fluid around a rotating  circular cylinder at Reynolds number $50\leq \mathsf{Re} \leq 500$

and dimensionless rotation rate $0\leq \alpha \leq 7$ has been carried out. The parametric

map of flow regimes has been constructed, where two zones of steady and two

zones of periodic solutions are signed out. The dependencies of the drag and

lift coefficients on the rotation rate for $\mathsf{Re}=200$ are studied in detail. Convergence

of the numerical solution to the known asymptotic solution at large $\alpha$ is confirmed.

}

\bigskip

{\normalsize\bf KEY WORDS:} {\it flow around the rotating cylinder, viscous fluid, vortex

shedding, Magnus effect, periodic flows, numerical simulation

}

\bigskip

\bigskip

\bigskip

\noindent{\bf$\mathbf1$. INTRODUCTION}

\bigskip

\noindent
Many scientific publications have been devoted to the problem of the flow of viscous
fluid around a rotating cylinder. The solution to this problem is determined by two
dimensionless parameters: Reynolds number $\mathsf{Re}=U_\infty d/\nu$ and rotation
rate $\alpha=\theta d/2 U_\infty$ (where $U_\infty$ is the oncoming free-stream velocity; $d$ is the
cylinder diameter; $\nu$ is kinematic viscosity of fluid; and $\theta$ is the angular velocity of the cylinder rotation).

It is well known, that the flow around a stationary cylinder at $\mathsf{Re}>47$ is characterized by the presence of the Karman vortex street in the wake of the cylinder  (see, e.g., Zdravkovich, 1997). The rotation of a circular cylinder
with constant rate $\alpha$ is accompanied by the origin of Magnus lift, which changes the structure
of the unsteady vortex wake and suppresses it at sufficiently large $\alpha$. One of the first experimental works
on this subject was presented by Prandtl $(1925)$, where it was shown that at $\alpha \ge \alpha_L \approx 2$ the
Karman vortex street is depressed and the flow pattern becomes steady. This conclusion conforms to numerous
experiments (Coutanceau and Menard, $1985$; Lam, $2009$) and numerical investigations (Kang et al., $1999$; Stojkovic
et al., $2002$; Mittal and Kumar, $2003$; Mazo and Morenko, $2006$; $Prikhod^,ko$ and Redtchits, $2009$). Based on his
investigations, Prandtl proposed the hypothesis that lift coefficient $c_l$ cannot be higher than $4\pi$.
The supposition that Magnus force is limited has not been confirmed (see, e.g., Tokumaro and Dimotakis, 1991).
Furthermore, it was shown by Moore $(1957)$, that in a two-dimensional case
at large rotation rates the drag coefficient $c_d$ and lift coefficient $c_l$ converge to the values
obtained in the context of the potential flow theory (Loytsanskiy, $1950$). It turned out that
$c_d \rightarrow 0$ at $\alpha \rightarrow \infty$, and the lift coefficient infinitely increases, namely, $c_l\rightarrow 2\pi\alpha$.
The results of calculations at $\mathsf{Re}=100$ and $\alpha\ge10$ agree well with this asymptotic, and the flow pattern obtained at these parameters is close to the potential flow pattern (Stojkovich et al., $2002$). However, it seems to be impossible to observe such a flow experimentally because of the significant influence of three-dimensional effects occurring at large rotational rates. Thus, Mittal ($2004$)
gave a comparison of computational results of a two- and three- dimensional problem of laminar flow around a rotating cylinder
at $\mathsf{Re}=200$ and $\alpha=5$, which qualitatively differed both in the flow pattern and the behavior of the integral flow characteristics.

The transition from steady solution at $\alpha\sim\alpha_L$ to an asymptotic one at at $\alpha\sim\alpha_H\gg\alpha_L$ is of special interest. The investigation of flow in the range of $\alpha_L\le\alpha\le\alpha_H$ is presented in few publications.
The supposition on the possibility of a periodic vortex descending at $\alpha\ge\alpha_L$ was first made by Chen et al. ($1993$) on the basis
of flow calculation at $\mathsf{Re}=200$ and $\alpha=3.25$. The latter investigations (Mittal and Kumar, $2003$) showed that at the given parameters the intensity of the descending vortex decays with time. However, in the range $4.34\le\alpha\le4.70$ the periodic solution is
indeed implemented, which differs from the known Karman vortex street. A similar result was obtained by Stojkovic et al. ($2002$) for the case of $\mathsf{Re}=100$ and $4.8\le\alpha\le5.15$.

This article is devoted to the detailed parametric investigation of steady and periodic regimes of laminar flow around rotational cylinder in a wide range of parameters $\mathsf{Re}$ and $\alpha$.

\bigskip

\bigskip

\noindent{\bf$\mathbf2$. THE MATHEMATICAL MODEL}

\bigskip

\noindent
Two dimensional incompressible flow was studied numerically using the Navier-Stokes equations formulated in terms of vorticity $\omega$ and stream function $\psi$:
\begin{equation}\label{psiom}
\frac{\partial{\omega}}{\partial{t}}+\vec v \nabla\omega=\frac{1}{\mathsf{Re}}\Delta\omega
\end{equation}

\begin{equation}\label{uv}
\begin{array}{cccc}
\Delta\psi=-\omega, & \omega=\displaystyle\frac{\mathstrut\partial{v_y}}{\partial{x}}-\frac{\partial{v_x}}{\partial{y}}, &
v_x=\displaystyle\frac{\mathstrut\partial{\psi}}{\partial{y}}, & v_y=\displaystyle-\frac{\mathstrut\partial{\psi}}{\partial{x}}
\end{array}
\end{equation}
\noindent
where $t$ is time; $x$ and $y$ are the Cartesian coordinates; $v_x$ and $v_y$ are the components of velocity vector $\vec v$;
$\mathsf{Re}=U_\infty d/\nu$; and $\nu$ is the kinematic viscosity.

\begin{figure}[h]
\begin{center}
\includegraphics[width=0.83\textwidth]{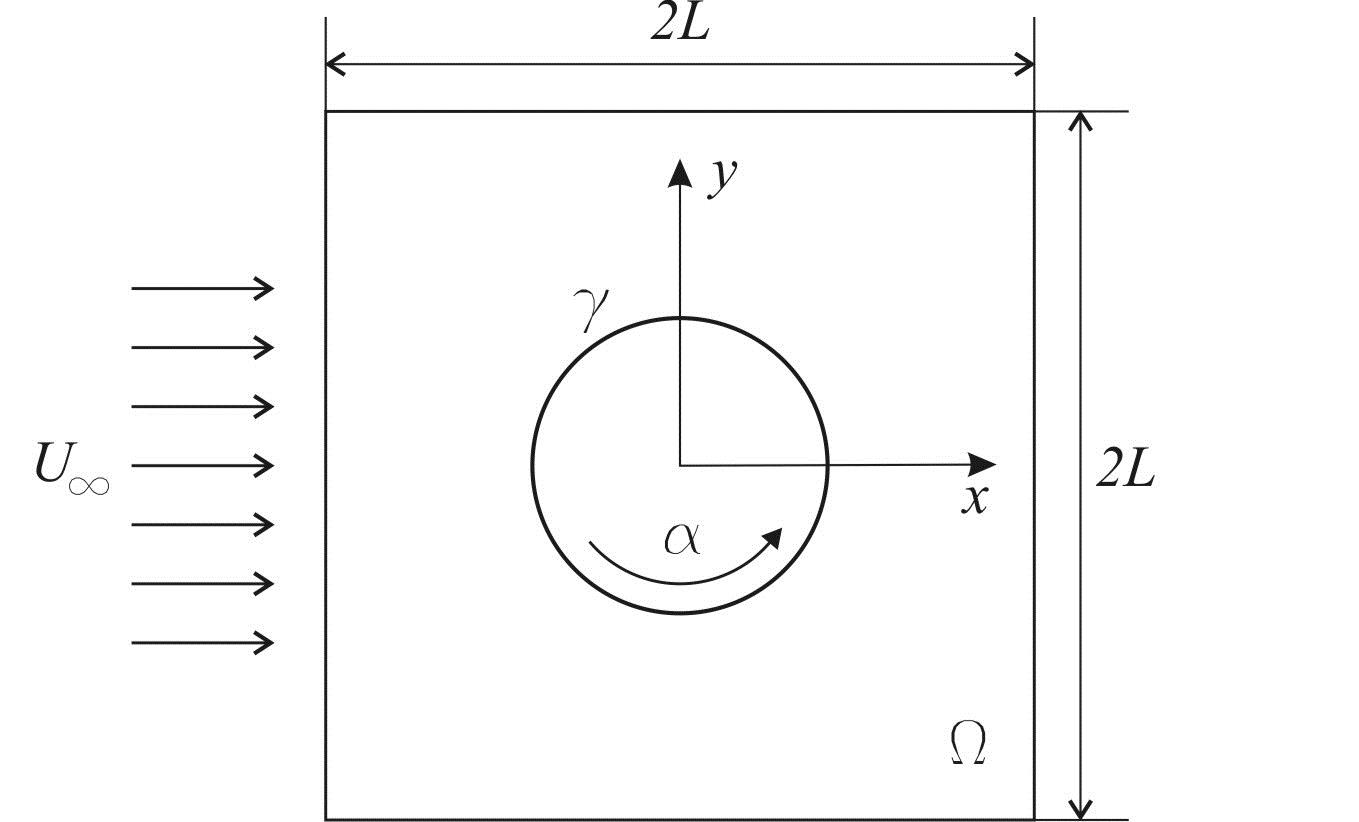}
\caption{Computational domain. \label{fluidarea}}
\end{center}
\end{figure}

Rectangular domain $\Omega=[-L,L]\times[-L,L]$ with a unit-diameter cylinder (Fig. \ref{fluidarea}) placed in its center was used to model infinite flow. The no slip condition was set on boundary $\gamma$ of streamlined body, accounting for cylinder

\noindent
counter-clockwise rotation with velocity $\alpha$:
\begin{equation}\label{noslip}
x,y\in\gamma : \psi=C(t),\;\frac{\partial{\psi}}{\partial{n}}=\alpha.
\end{equation}
\noindent
where $n$ is the outer normal to $\gamma$. To determine function $C(t)$ the approach suggested by Glowinski ($2003$) was used, which is based on supplementation of the problem statement with non-local boundary conditions
\begin{equation}\label{Pirson}
\int\limits_{\gamma}\left(\frac{\partial{v_s}}{\partial{t}}-\frac{1}{\mathsf{Re}}\frac{\partial{\omega}}{\partial{n}}\right)d\gamma=0.
\end{equation}
\noindent
Condition (\ref{Pirson}) can be obtained using integration of Pierson relations (Fletcher, $1991$):
\begin{equation}\notag
x,y\in\gamma : \frac{\partial{v_s}}{\partial{t}}=\frac{1}{\mathsf{Re}}\Delta v_s-\frac{\partial{p}}{\partial{s}}
\end{equation}
\noindent
where $s$ is the tangent line to contour $\gamma$ and $p$ is pressure.

The impulsive start of rotation and translation motion of the cylinder from quiescent state
$\psi=0$, $\omega=0$ is modeled. The tangential velocity of fluid on contour $\gamma$ is
given by formula $v_s=-h(t)\alpha$, and boundary condition (\ref{Pirson}) becomes:
\begin{equation}\label{Pirsonmod}
\int\limits_{\gamma}\frac{\partial{\omega}}{\partial{n}}d\gamma=-\mathsf{Re}\pi\alpha\delta(t)
\end{equation}
\noindent
where $h$ is the Heaviside function and $\delta$ is the delta function.

In the inlet section, the potential flow of fluid with rate $Q=2L$ is specified in the absence
of the transversal velocity component:
\begin{equation}\label{incond1}
x=-L: \frac{\partial{\psi}}{\partial{n}}=0, \omega=0
\end{equation}
\begin{equation}\label{incond2}
y=\pm L:\psi=y, \omega=0
\end{equation}

The latter condition in Eq.(\ref{incond2}) simulates the condition of ideal slip on the horizontal
boundaries of the computational domain.

In the outlet section $x=L$, the convective boundary conditions are set (Orlanski, $1976$):
\begin{equation}\label{outcond}
x=L: \frac{\partial{\omega}}{\partial{t}}+\vec u \cdot\nabla\omega=0, \frac{\partial{\psi}}{\partial{t}}+\vec u \cdot\nabla\omega=0
\end{equation}
\noindent
where $\vec u=(1,0)$ is the flow-average velocity. Conditions (\ref{incond1})-(\ref{outcond}) provide the absence of a
noticeable influence of the boundary conditions on the numerical solution at the sufficient distance of the computational
area boundaries.

Hydrodynamic pressure forces $R^p$ and friction forces $R^f$ and the corresponding integral drag and lift coefficients
are calculated using the following formulas:
\begin{equation}\label{clcd}
\begin{array}{cll}
R^p(l)=2p(l), R^f(l)=\displaystyle\frac{1}{\mathsf{Re}}\omega(l)
\\
c^p_d=\displaystyle\int\limits_{\gamma}R^p n_xd\gamma, c^f_d=\displaystyle\int\limits_{\gamma}R^f n_yd\gamma, c_d=c^p_d+c^f_d
\\
c^p_l=\displaystyle\int\limits_{\gamma}R^p n_yd\gamma, c^f_l=\displaystyle\int\limits_{\gamma}R^f n_xd\gamma, c_l=c^p_l+c^f_l
\end{array}
\end{equation}

Here, $l$ is the lenght of the arc, counting from the front point of the cylinder; and $n_x$ and $n_y$ are cosines of the angles between the normal to $\gamma$ and the corresponding axes. The boundary problem for Bernoulli variable $B=p+v^2/2$ is solved to determine the pressure $p$ distribution over contour $\gamma$, which is involved in relation (\ref{clcd}) (see Fletcher, $1991$; Mazo and Morenko, $2006$); namely:

\begin{equation}\label{Ber}
\begin{array}{cl}
-\Delta B=\nabla\psi\cdot\nabla\omega-\omega^2
\\
x,y\in\gamma:\displaystyle\frac{\partial{B}}{\partial{n}}=-\frac{1}{\mathsf{Re}}\frac{\partial{\omega}}{\partial{s}}+v_s\omega, x,y=\pm L: \frac{\partial{B}}{\partial{n}}=0
\end{array}
\end{equation}
\noindent
To determine the unique solution of degenerate problem (\ref{Ber}), pressure $p=0$ $(B=0.5)$ is set at single point of the inlet section.

\bigskip

\bigskip

\noindent{\bf$\mathbf3$. SOLUTION TECHNIQUE}

\bigskip

\noindent
The main difficulty in solving the set of Eqs.(\ref{psiom}) and (\ref{uv}) with boundary conditions (\ref{noslip}) and (\ref{Pirsonmod})-(\ref{outcond}) consists in the presence of non-local relation (\ref{Pirsonmod}) in the problem statement,
which is used for calculation of value $C$ of the stream function $\gamma$. To express the unknown hydrodynamic functions
in terms of $C$, a double-layer linearized scheme with time-step $\tau$ is set for Eq.(\ref{psiom}) of the vorticity transfer;
therefore, we have:
\begin{equation}\label{linscheme}
\frac{\omega-\breve\omega}{\tau}-\frac{1}{\mathsf{Re}}\Delta\omega=-\breve{\vec{v}}\cdot\nabla\breve\omega
\end{equation}
\noindent
where $\breve\omega(t)=\omega(t-\tau)$.

Function $C$ is constant on every time layer $t$, and the set of differential equations (\ref{uv}) and (\ref{linscheme}) is linear
with respect to $\psi,\omega$. Therefore, the unknown functions linearly depend on constant $C$:
\begin{equation}\label{line}
\omega=\xi\omega_{1}+(1-\xi)\omega_{2}, \psi=\xi\psi_{1}+(1-\xi)\psi_{2}, C=\xi C_1+(1-\xi)C_2
\end{equation}
\noindent
where $\psi_1$, $\omega_1$ and $\psi_2$, $\omega_2$ are the solutions of problems (\ref{uv}) and (\ref{linscheme}) with boundary
conditions (\ref{noslip}) and (\ref{incond1})-(\ref{outcond}) for two fixed values $C_1$, $C_2$. To determine constant $\xi$ (or $C$)
it is sufficient to substitute solution (\ref{line}) in non-local boundary condition (\ref{Pirsonmod}) and solve the following linear equation:
\begin{equation}\notag
\xi\int\limits_{\gamma}\frac{\partial{\omega_1}}{\partial{n}}d\gamma+(1-\xi)\int\limits_{\gamma}\frac{\partial{\omega_2}}{\partial{n}}d\gamma=-\mathsf{Re}\pi\alpha\delta(t)
\end{equation}

Furthermore, we will briefly describe the iterative process of the solution of the set of Eqs.(\ref{uv}) and (\ref{linscheme})
with boundary conditions (\ref{noslip}) and (\ref{incond1})-(\ref{outcond}) at defined value $C$.

Before the iterations start $\psi$, $\omega$ in the outlet sections are found using approximation of boundary conditions (\ref{outcond})
as follows:

\begin{equation}\label{outdis}
x=L: \omega=\breve\omega - \tau\frac{\partial{\breve\omega}}{\partial{x}}, \psi=\psi=\breve\psi - \tau\frac{\partial{\breve\psi}}{\partial{x}}
\end{equation}

Functions $\psi$, $\omega$ from the previous time layer are used as the initial approximation of the iterative process.

In the $k$ step of the iterative process the flow function is determined as the Poisson equation $[Eq.(\ref{uv})]$ solution with boundary
conditions (\ref{incond1}), (\ref{incond2}), and (\ref{outdis}) and value $\omega$ from previous iteration on the right-hand side:
\begin{equation}\label{iter}
-\Delta\psi^k=\omega^{k-1},\hspace{0.2cm}x,y\in\gamma: \psi^k=C
\end{equation}

Furthermore, the boundary values of vorticity $\omega_b$ on contour $\gamma$ are defined using Neumann conditions (\ref{noslip})
(see Mazo and Dautov, $2005$):
\begin{equation}\label{Neumann}
x,y\in\gamma: \omega^{k}_{b}=-\Delta \psi^k,\hspace{0.2cm} \frac{\partial{\psi^k}}{\partial{n}}=\alpha
\end{equation}

Next, the equation of vorticity transfer(\ref{linscheme}) is solved with boundary conditions (\ref{incond1}), (\ref{incond2}), (\ref{outdis}), and (\ref{Neumann}). To speed up the iterative process
convergence of boundary values (\ref{Neumann}) are applied:

\begin{equation}\label{conv}
x,y\in\gamma: \omega^{k}=\zeta\omega^{k}_{b}+(1-\zeta)\omega^{k-1}_{b}
\end{equation}

The criterion of iteration exit is taken in the following form:

\begin{equation}\label{cond}
\max_\gamma\left|\displaystyle\frac{\partial{\psi^k}}{\partial{n}}-\alpha\right|<\varepsilon
\end{equation}

With the relaxation coeficient $\zeta\approx0.5$ the realization of criterion (\ref{cond}) at $\varepsilon=10^{-6}$ is reached
in two or three steps.

The spatial approximation of differential Eqs.(\ref{linscheme}),(\ref{iter}), and (\ref{Neumann}) is performed using the finite-element method (Fletcher, $1991$; Kalinin and Mazo, $2009$) on an unstructured grid of bilinear elements. Approximation of the convective term is carried out using the TVD approach (Kuzmin and Turek, $2004$) with Superbee limiter. Derivatives $\partial{\psi}/\partial{n}$, $\partial{\omega}/\partial{n}$ in relationships
(\ref{Pirsonmod}), (\ref{clcd}), and (\ref{cond}) are calculated using the technique described by Kalinin and Mazo ($2009$). The sets of linear equations obtained as a result of the finite-element approximation are solved using the algebraic multigrid method (Trottenberg et al., $2001$).

The scheme obtained has the first time order of approximation and second spatial order of approximation.

\bigskip

\bigskip

\noindent{\bf$\mathbf{3.1}$. Results of the Calculation}

\bigskip

\noindent
Numerical simulation of laminar flow around a rotating cylinder in the range of parameters $50\leq\mathsf{Re}\leq500$
and $0\leq\alpha\leq7$ were carried out. As a result of a series of tests it was revealed that, for these values, it is sufficient ti use a grid with a spatial step of $0.002$ near contour $\gamma$ and time step $\tau=0.001$; when choosing $L\geq100$, the influence of the outlet
boundary conditions on the problem solution becomes negligible.

The calculations showed that both steady and periodic flow regimes were observed depending on rotation rate $\alpha$. Some results of simulation for $\mathsf{Re}=200$ are presented below.

\bigskip

\bigskip

\noindent{\bf$\mathbf{3.2}$. The Periodic Flow at Small Rotational Velocities}

\bigskip

\begin{figure}[h]
\begin{center}
\includegraphics[width=0.83\textwidth]{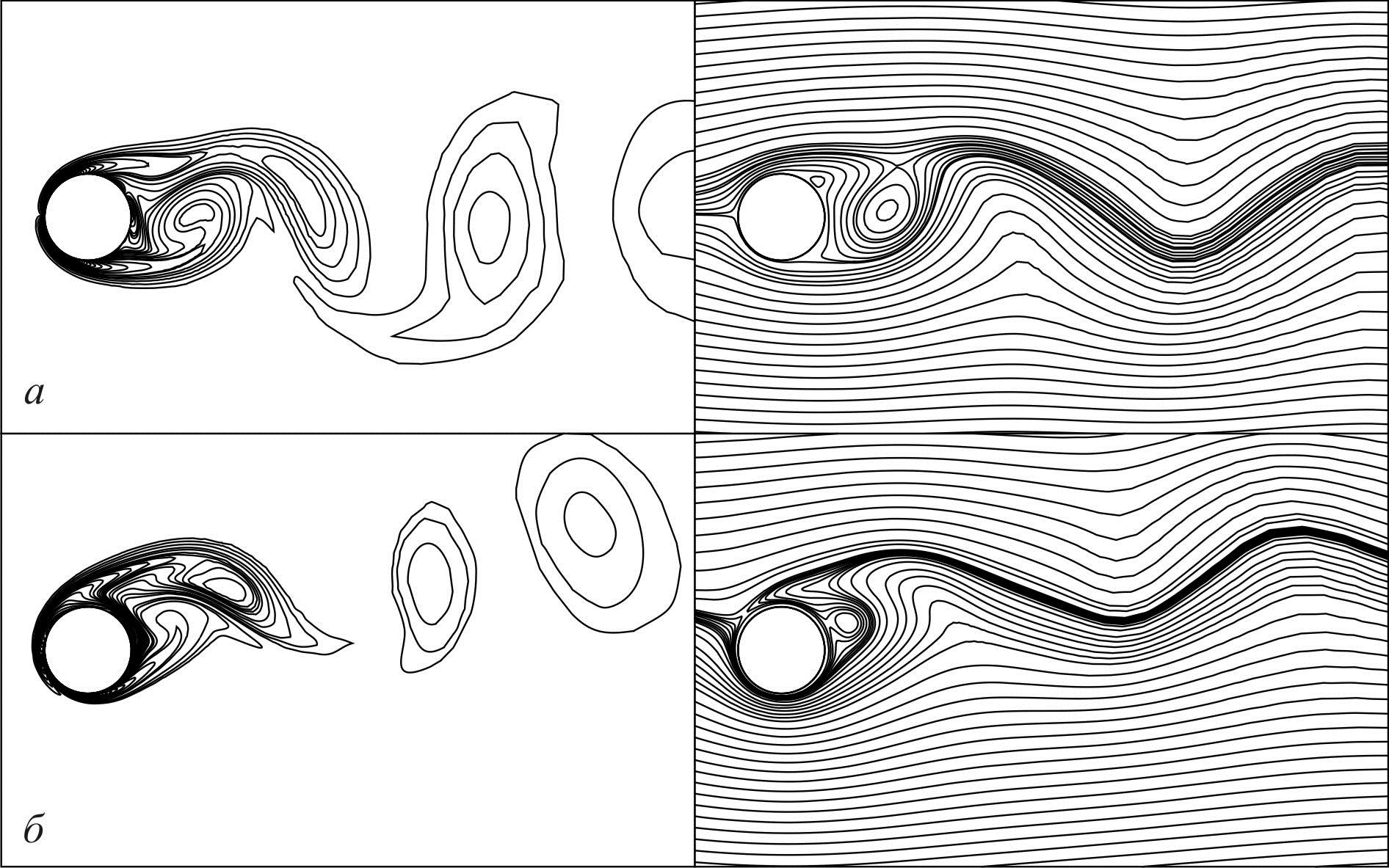}
\caption{Instantaneous vorticity field (on the left) and streamlines (ont the right) in the case of periodic flow around the cylinder at $\mathsf{Re}=200$ when rotating at (a) $\alpha=0$ and (b) $\alpha=1.5$. \label{vort}}
\end{center}
\end{figure}

\noindent
The periodic flow around a fixed ($\alpha=0$) cylinder at $\mathsf{Re}>47$, accompanied by the vortex separation from the upper and lower cylinder cheeks and formation of the Karman street [Fig. \ref{vort}(a)], was investigated and has been described in detail in the literature.
The main characteristics of this flow are given below for the sake of further comparison with the cases of positive $\alpha$.

\begin{figure}[h]
\begin{center}
\includegraphics[width=0.83\textwidth]{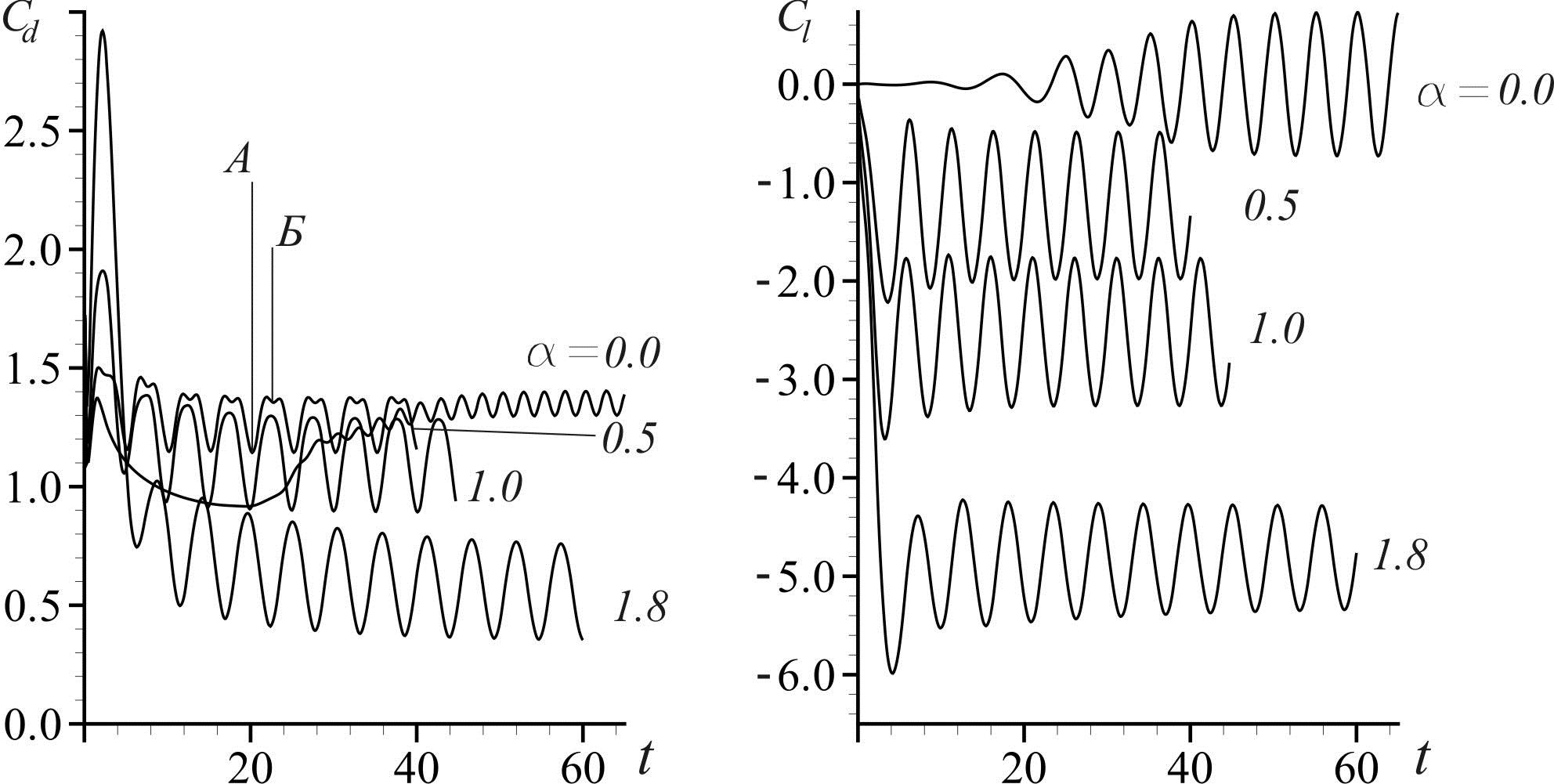}
\caption{Variations of drag coefficient $c_d$ (on the left) and lift coefficient (on the right) in time at $\mathsf{Re}=200$ at small rates of rotation $\alpha$.\label{drag}}
\end{center}
\end{figure}

Generation of the lower vortex with positive vorticity accompanied by the appearance of a stagnation zone near the lower cheek and the bottom
part of the near wake that led to a pressure increase in this zone; as a consequence, the lift and drag coefficients increased. Generation of the upper vortex with negative $\omega$ also was accompanied by an increase of $c_d$, although $c_l$ was diminished. Thus, the values of $c_l$ and $c_d$ performed sinusoidal oscillations in time (Fig. \ref{drag}, see curves $\alpha=0$); the frequency of oscillations $c_d$ were twice as much as the frequency of oscillations $c_l$.

The rotation of the cylinder disturbed the flow symmetry. The upper vortex, being shifted in the rotation direction in the case of $\alpha=0$. more intensively stagnated the external flow, whereas the lower vortex stagnated the flow less intensively because the flow was "hidden" behind the cylinder [Fig. \ref{vort}(b)]. Therefore, the upper vortex separation was determinative in the $c_d$ self-oscillations. This effect was illustrated by the distribution of local minima of curve $\alpha=0.5$ in Fig. \ref{drag} (where separation of the upper vortex corresponds to moment $A$ and the separation of the lower one corresponds to moment $B$). Furthermore, at $\alpha>0.6$ the vortex separation from below did not lead to the appearance of local minima of function $c_d(t)$. At the same time, the drag coefficient again varied under the sinusoidal flow; however, unlike the flow around stationary cylinder its frequency was equal to oscillation frequency $c_l$ (Fig. \ref{drag}, see curves $\alpha=1$ and $1.8$).

\bigskip

\bigskip

\noindent{\bf$\mathbf{3.3}$. The Steady Flow at Moderate Rotational Velocities}

\bigskip

\noindent
The periodic regime of flow around the cylinder was implemented at the rotation rate, which was less than critical value $\alpha\le\alpha_L$. In previous works (Kang et al., $1999$; Mazo and Morenko, $2006$; $Prikhod^,ko$ and Redtchits, $2009$), and many others, it has been shown that at $\mathsf{Re}>100$ the value $\alpha_L$ weakly depends on the choice of Reynolds number and is approximately equal to $2$. Our calculations confirm these conclusions. So, at $\mathsf{Re}>200$ critical rotation rate equals $\alpha_L=1.9$.

\begin{figure}[h]
\begin{center}
\includegraphics[width=0.83\textwidth]{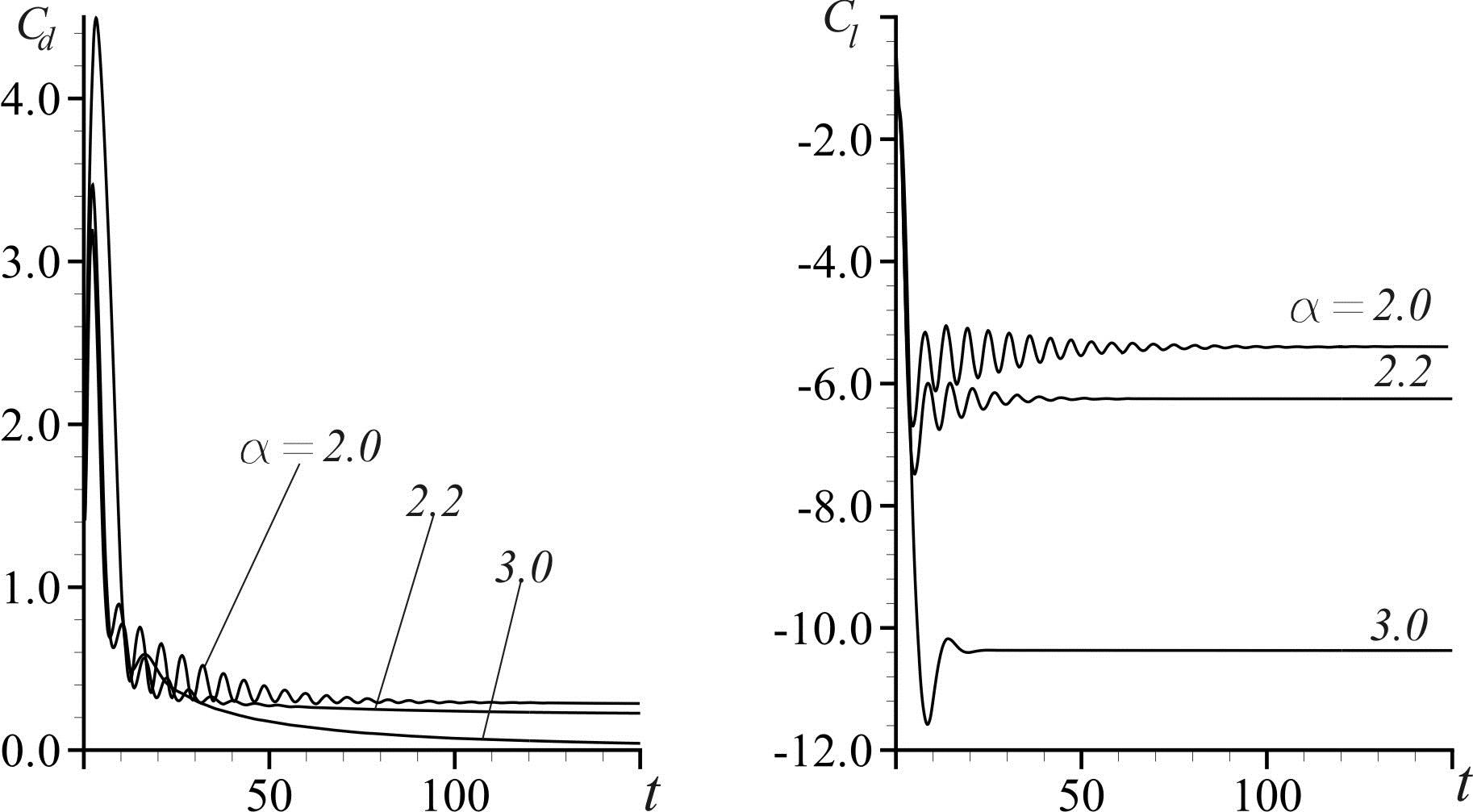}
\caption{Variation of drag coefficient $c_d$ (on the left) and lift coefficient (on the right) in time at $\mathsf{Re}=200$ at moderate rates of rotation $\alpha$.\label{drag1}}
\end{center}
\end{figure}

If $\alpha$ insignificantly exceeds $\alpha_L$, then the Karman street behind the cylinder forms right after the instantaneous start of rotation; however, it is oppressed with time. Thus, the number of vortices separated from the cylinder ar this regime is finite. The vortex suppression in the wake was accompanied by decay of oscillations of the drag and lift coefficients (Fig. \ref{drag1}, curves $\alpha=2$ and $2.2$). The corresponding flow

\begin{figure}[h]
\begin{center}
\includegraphics[width=0.83\textwidth]{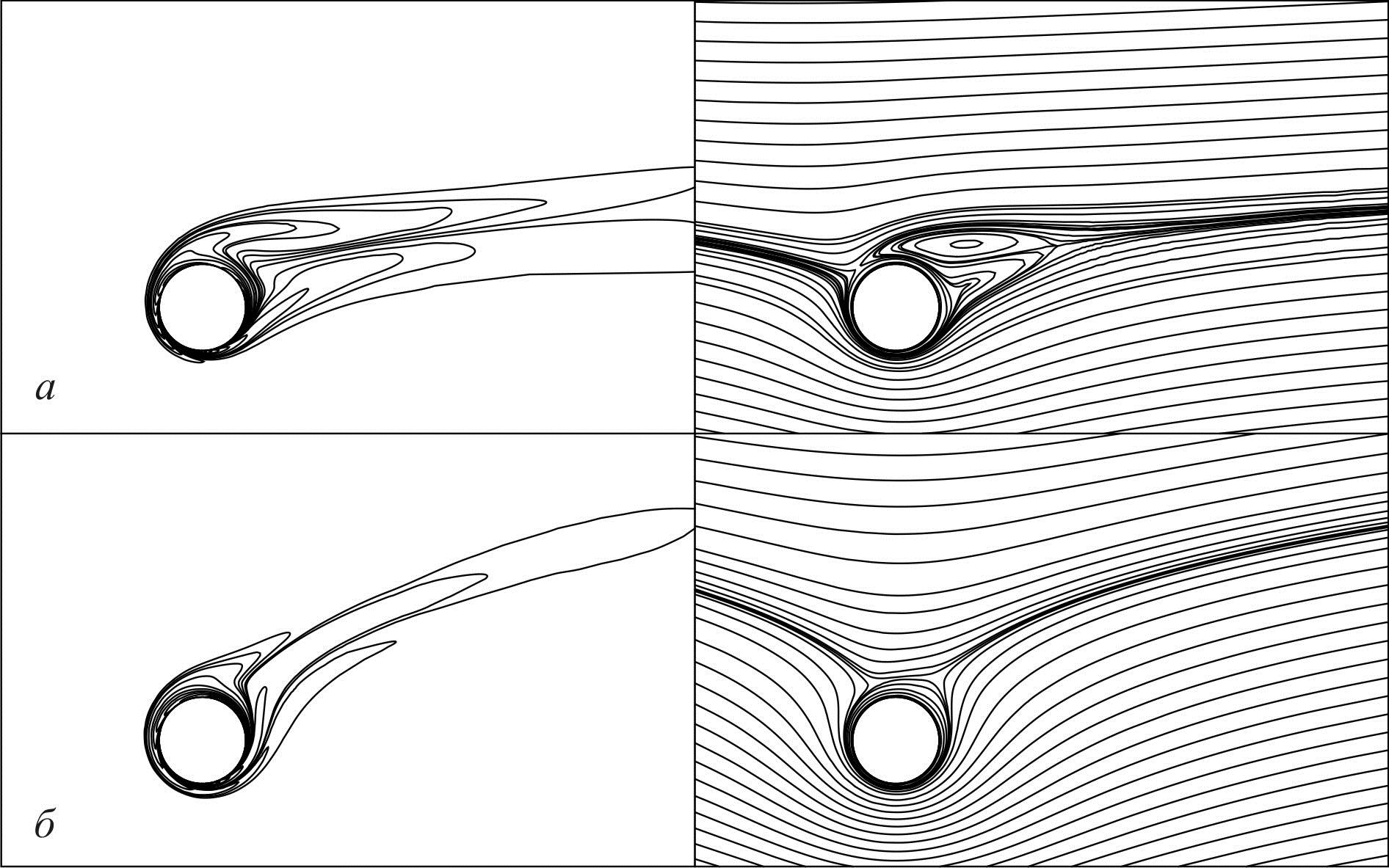}
\caption{Vorticity field (on the left) and streamlines (on the right) in the case of steady flow around the cylinder at $\mathsf{Re}=200$ when rotating at (a) $\alpha=2$ and (b) $\alpha=3.4$. \label{vort1}}
\end{center}
\end{figure}

\noindent
pattern, shown in Fig. \ref{vort1}(a), is characterized by the presence of a zone with closed stream-lines.

As the rotation rate increases, the number of vortices descending from the cylinder before the moment of stabilization decreases; and, simultaneously, the dimension of the zone with closed streamlines decreases. At $\alpha>2.7$ only the start vortex descends from the cylinder. This may be seen from the monotonous character of the stabilization of values $c_d$ and $c_l$ starting from $t=25$ (Fig. \ref{drag1}, curves $\alpha=3$). At $\alpha>3.2$ the zone of closed flow also vanishes [Fig. \ref{vort1}(b)]. At the same time, only one stagnation point forms near the cylinder; namely, the point of intersection of streamlines, which is typical for the potential solution of the problem at $\alpha>2$.

\bigskip

\bigskip

\bigskip

\noindent{\bf$\mathbf{3.4}$. Periodic Flow at Large Rotational Rates}

\bigskip

\begin{figure}[h]
\begin{center}
\includegraphics[width=0.68\textwidth]{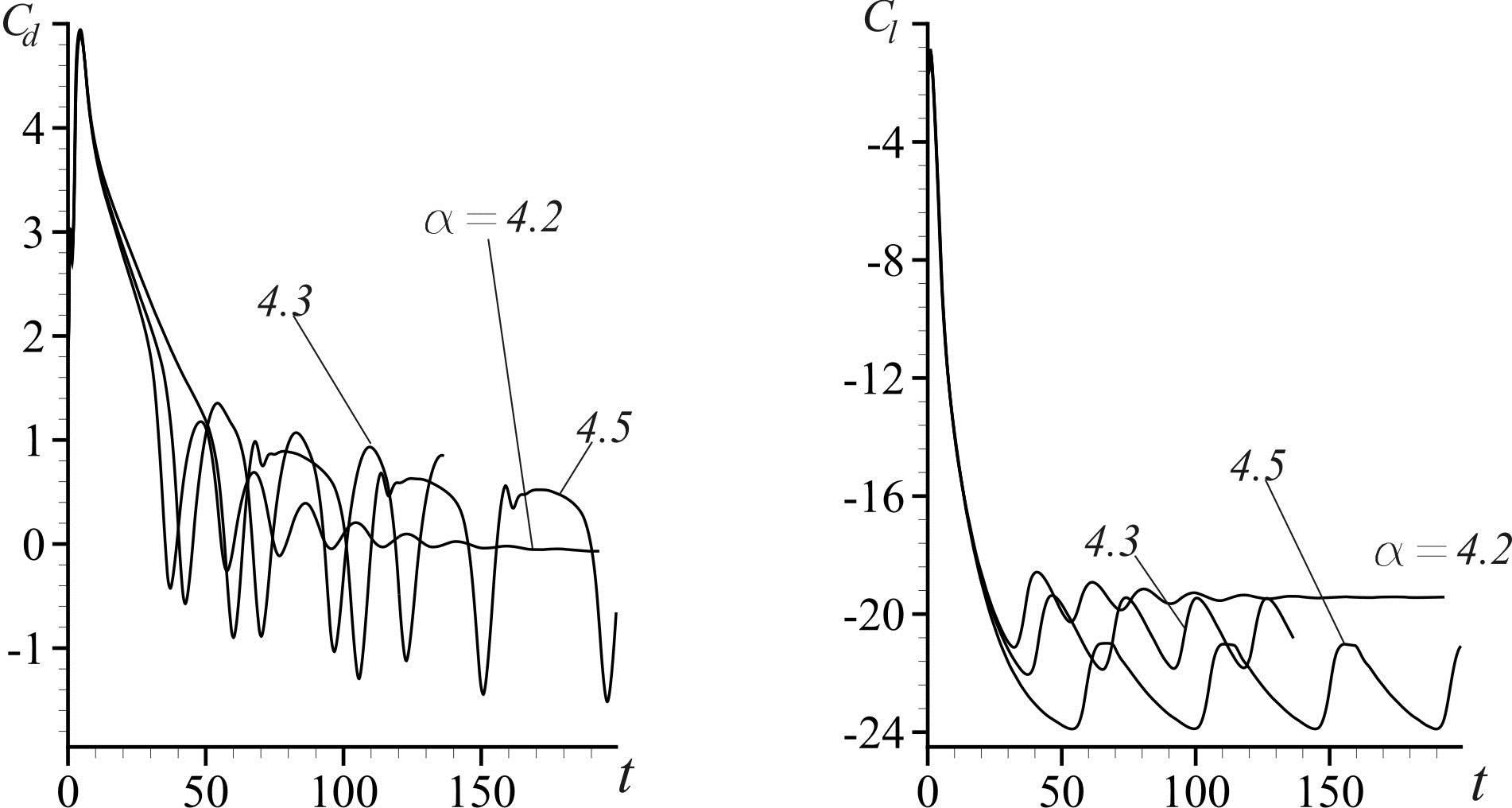}
\caption{Variation of drag coefficient $c_d$ (on the left) and lift coefficient $c_l$ (on the right) at $\mathsf{Re}=200$ in time in the case of periodic regimes for large rates of rotation $\alpha$. \label{drag2}}
\end{center}
\end{figure}

\noindent
With further increase of the rotational rate ($\alpha>3.8$),stabilization of coefficients $c_d$ and $c_l$
is again accompanied by decaying oscillations (Fig. \ref{drag2}, see curves $\alpha=4.2$), which indicates descending of several vortices from the moment of rotation start up to attainment of the steady flow regime. With the increase of $\alpha$ the amount of descended vortices increases, and in the range of $4.3\leq\alpha\leq4.6$ periodic flow is realized.

\begin{figure}[h]
\begin{center}
\includegraphics[width=0.68\textwidth]{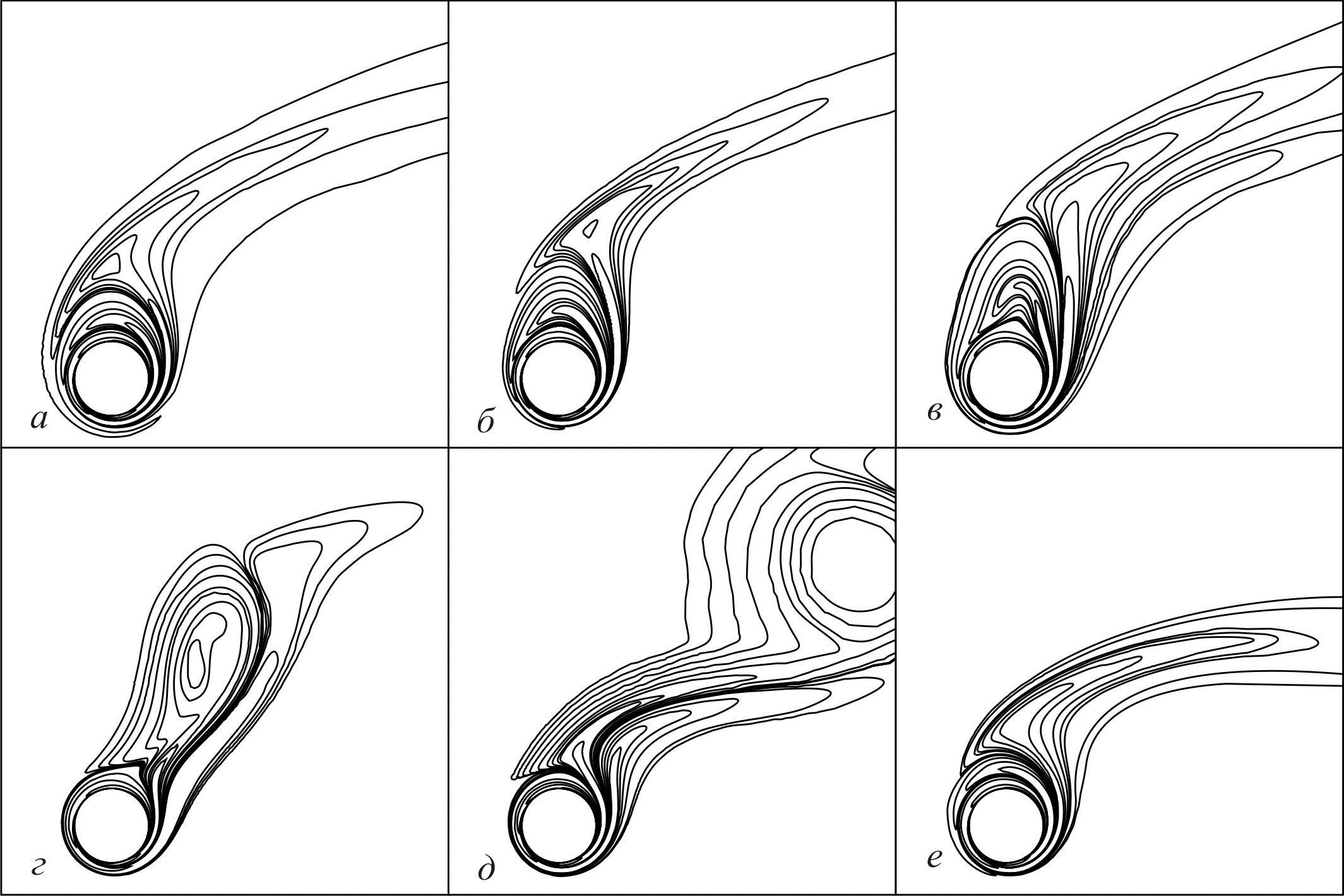}
\caption{Vortex separation at periodic flow regime $\mathsf{Re}=200$ and $\alpha=4.5$. \label{vortsep}}
\end{center}
\end{figure}

In this case, the oscillations of $c_d$ and $c_l$ are not harmonic, and their amplitude and period are substantially greater than in the Karman street at $\alpha<\alpha_L$ (Fig. \ref{drag2}, see curves $\alpha=4.3$ and $4.5$). The flow pattern also differs from the flow pattern shown in Fig. \ref{vort}(b). At large rates of rotation the entire vortex structure in the near wake shifts to the upper cheek of the cylinder, and vortex separations also accur there. The periodic process,

\noindent
shown in Fig. \ref{vortsep}, consist of a gradual accumulation of positive vorticity above the cylinder [Figs. \ref{vortsep}(a)-(c)] and its separations [Figs. \ref{vortsep}(d)-(f)], accompanied by a drastic decrease of the drag coefficient and simultaneous increase of lift (Fig.\ref{drag2}, curves $\alpha=4.3$ and $4.5$).

\begin{figure}[h]
\begin{center}
\includegraphics[width=0.68\textwidth]{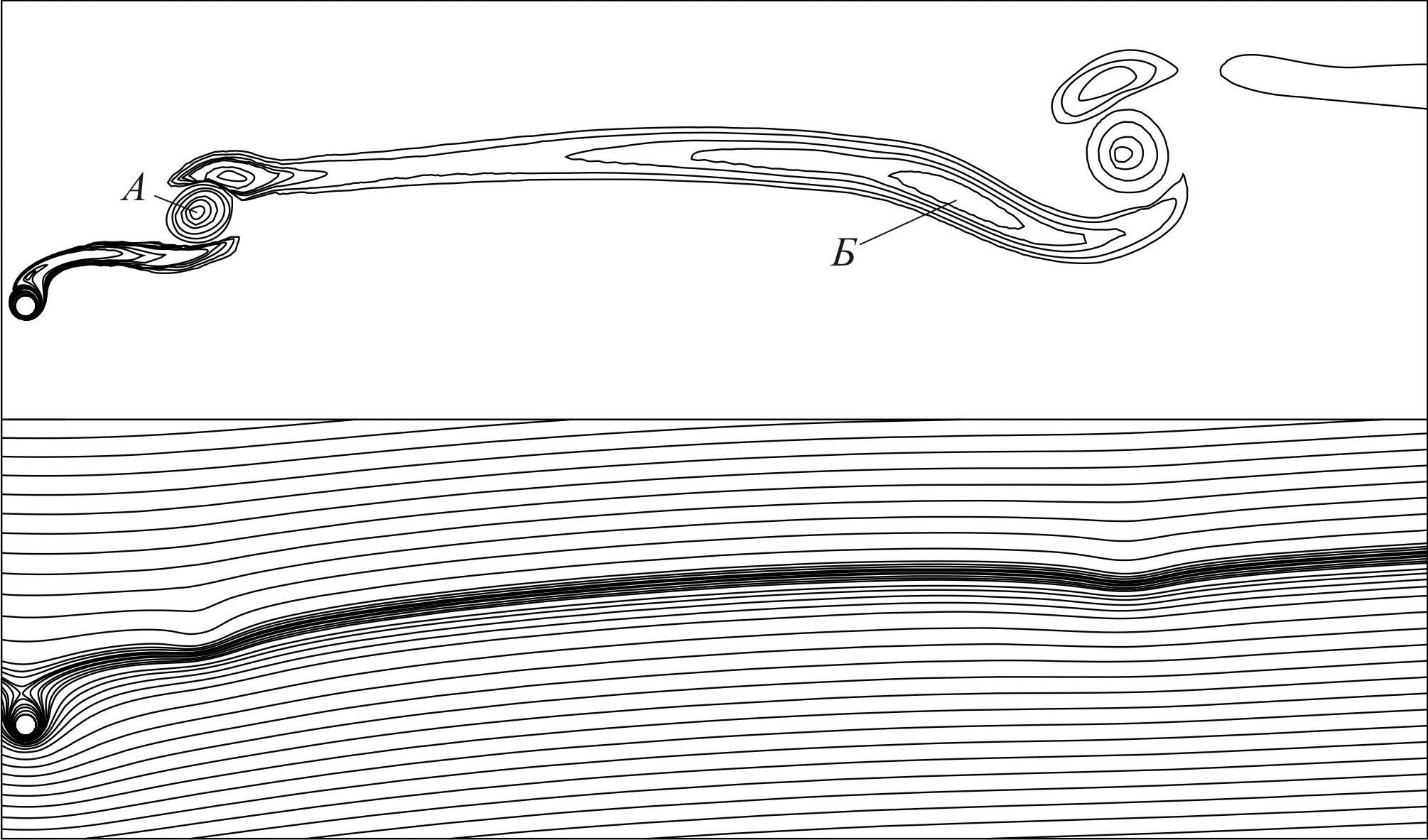}
\caption{Instantaneous vorticity field (on top) and streamlines (below) in the case of periodic flow at $\mathsf{Re}=200$ and $\alpha=4.5$. The zones are designated as follows: (A) $\omega>0$ and (B) $\omega<0$. \label{tracer}}
\end{center}
\end{figure}

\begin{figure}[h]
\begin{center}
\includegraphics[width=0.68\textwidth]{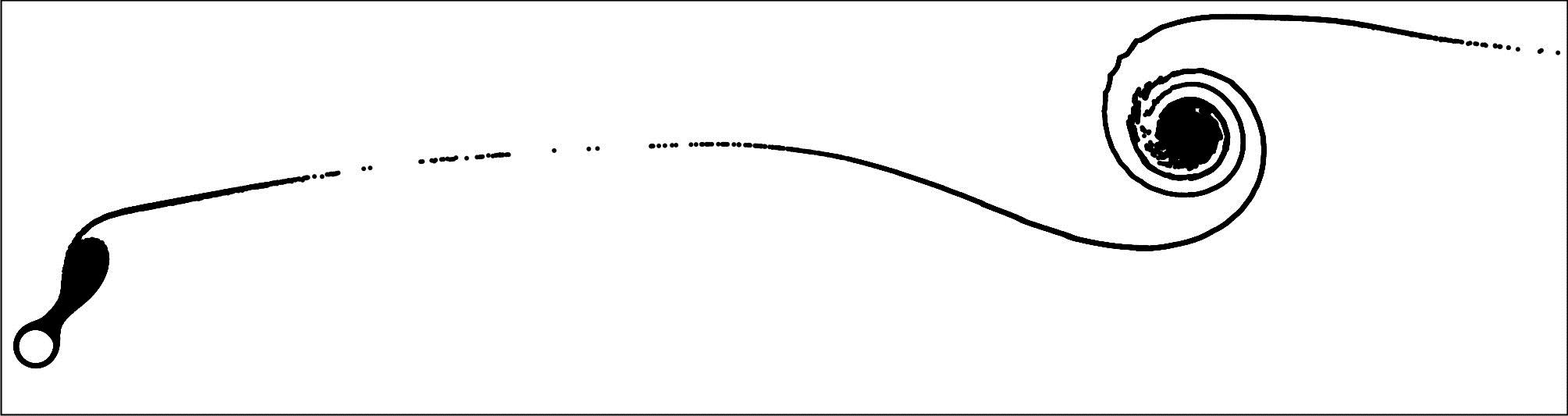}
\caption{Vortex separation for flow at $\mathsf{Re}=200$ and $\alpha=4.5$, visualized by means of tracers. \label{tracer1}}
\end{center}
\end{figure}

In the far wake behind the streamlined body (Fig. \ref{tracer}) the positive vorticity is concentrated in circular areas $A$, arising as a result of fast separation, and the negative vorticity is spread out inside prolonged vortex zones $B$, which are formed over a long period of time due to positive vortex generation. In Fig. \ref{tracer1} this periodic flow regime is visualized using tracers outflowing from the cylinder surface.

\newpage
\bigskip

\bigskip

\noindent{\bf$\mathbf{3.5}$. The Steady Solution at Large Rates of Rotations}

\bigskip

\begin{figure}[h]
\begin{center}
\includegraphics[width=0.68\textwidth]{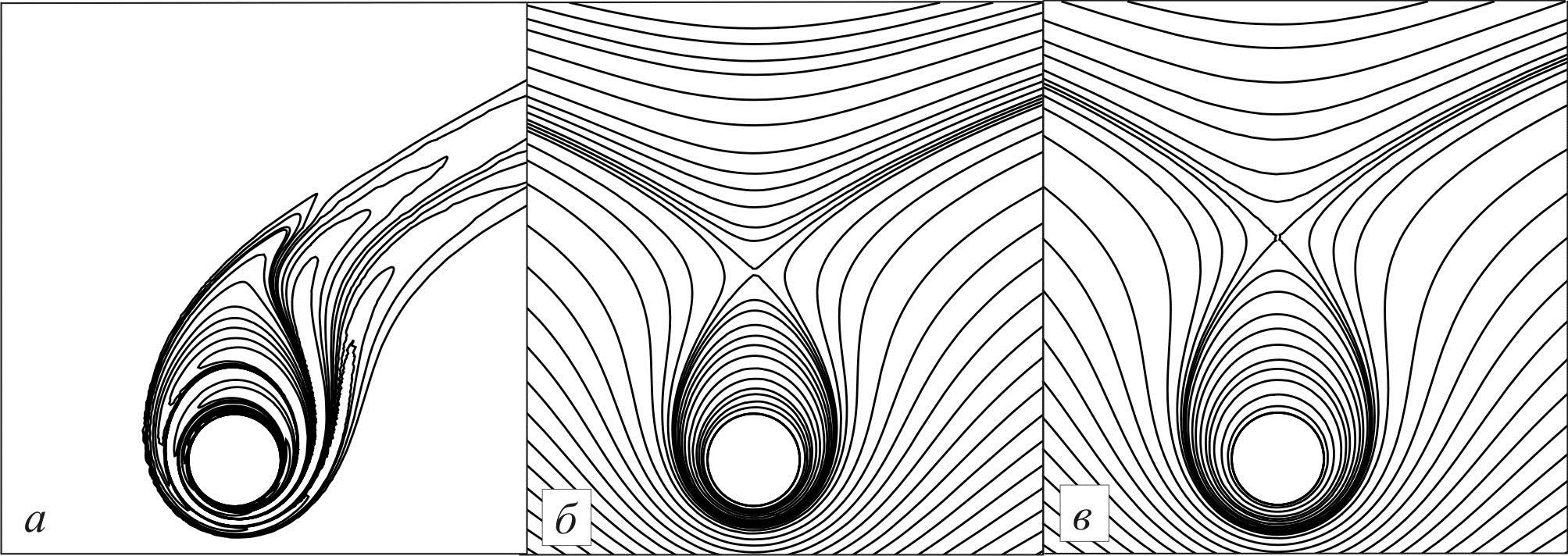}
\caption{Vorticity field (a) and streamlines (b) in the case of steady flow around the cylinder at $\mathsf{Re}=200$ when rotating with $\alpha=4.5$; Streamlines in the case of potential flow at $\alpha=5$ (c).\label{vf}}
\end{center}
\end{figure}

\begin{figure}[h]
\begin{center}
\includegraphics[width=0.68\textwidth]{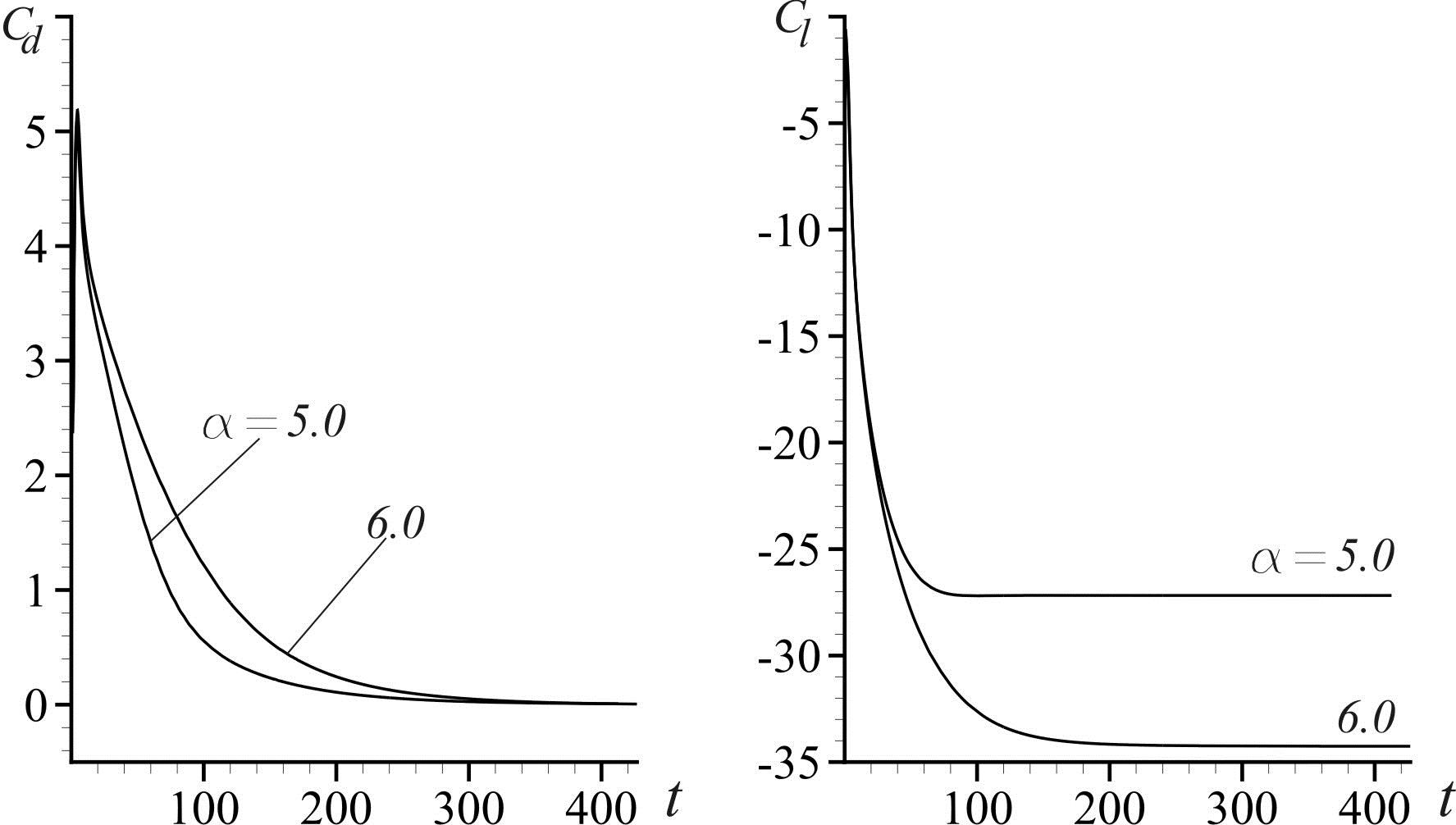}
\caption{Variation in time of drag coefficient $c_d$ (on the left) and lift coefficient $c_l$ (on the right) at $\mathsf{Re}=200$ in the case of steady regimes for large rates of rotation $\alpha$.\label{vt}}
\end{center}
\end{figure}

\noindent
With large rates of rotation $\alpha>\alpha_H\approx4.6$ the problem solution again provides a steady flow pattern, shown in Figs. \ref{vf}(a) and (b). The egg-shaped closed streamlines with a stagnation point located strictly above the cylinder are formed around the cylinder. The streamlines of this flow are close to the known solution of problem (see Loytsansky, $1950$) within the potential theory [Fig. \ref{vf}(c)].
The drag coefficient appears to be equal to zero, which coincides with asymptotic behaviour for large rates of rotation (Moore, $1957$).
The stabilization of coefficients $c_d$ and $c_l$ after the instantaneous rotation start and the vortex descending start is of monotonous character (Fig. \ref{vt}); the time of stabilization

\vspace{0.5cm}

\noindent
substantially exceeds the time of attainment of the steady regime at moderate values of $\alpha$ and increases with the increase of rotation rate.

\begin{figure}[h]
\begin{center}
\includegraphics[width=0.68\textwidth]{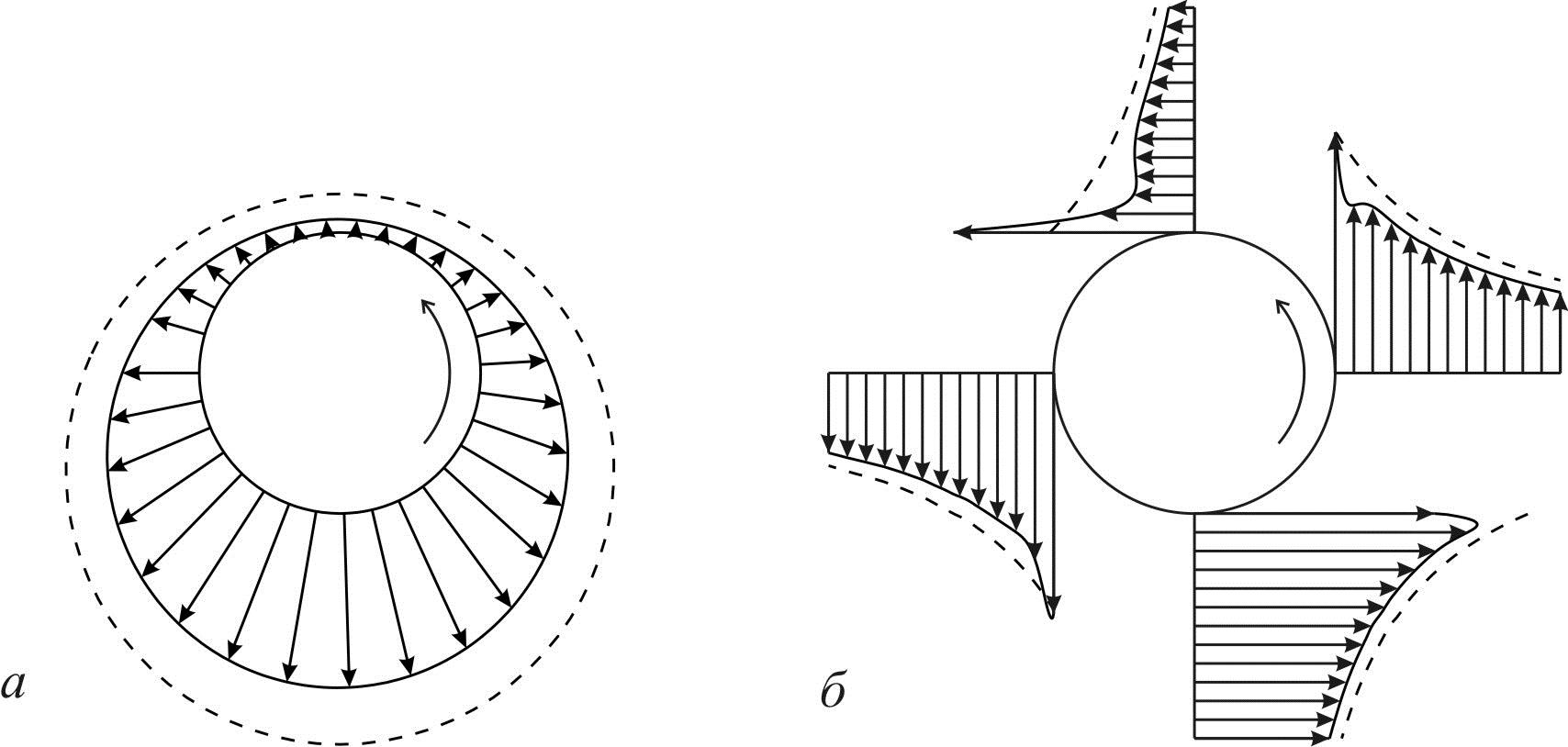}
\caption{Distribution of pressure forces (a) and velocity diagram (b) along the cylinder contour at $\alpha=5:$  --- corresponds to the result of the calculation at $\mathsf{Re}=200$, - - - corresponds to the potential solution.\label{pressd}}
\end{center}
\end{figure}

In Fig. \ref{pressd}(a), the $\gamma$ distribution of $R^p$ is given. Art high rates of rotation the pressure is negative along the entire streamlined contour (let us remember that the pressure in the inlet section is taken to be zero). At the same time the absolute values of $R^p$ near the the upper one; therefore, the resultant force of pressure forces is directed downward. The obtained pattern of the $R^p$ distribution for viscous flow qualitatively agrees with the potential solution, although it sufficiently differs in quantity. The diagrams of velocity on the cylinder boundary are also qualitatively similar [Fig. \ref{pressd}(b)]. Near the lower cheek, on the contrary, the rotational and the incoming flow velocities are added, the fluid velocity decreases less intensively, and in some area inside the boundary layer it even exceeds the rotation rate of $\alpha=5$.

\bigskip

\noindent{\bf$\mathbf{3.6}$. The Drag and Lift Coefficients}

\bigskip

\begin{figure}[h]
\begin{center}
\includegraphics[width=0.58\textwidth]{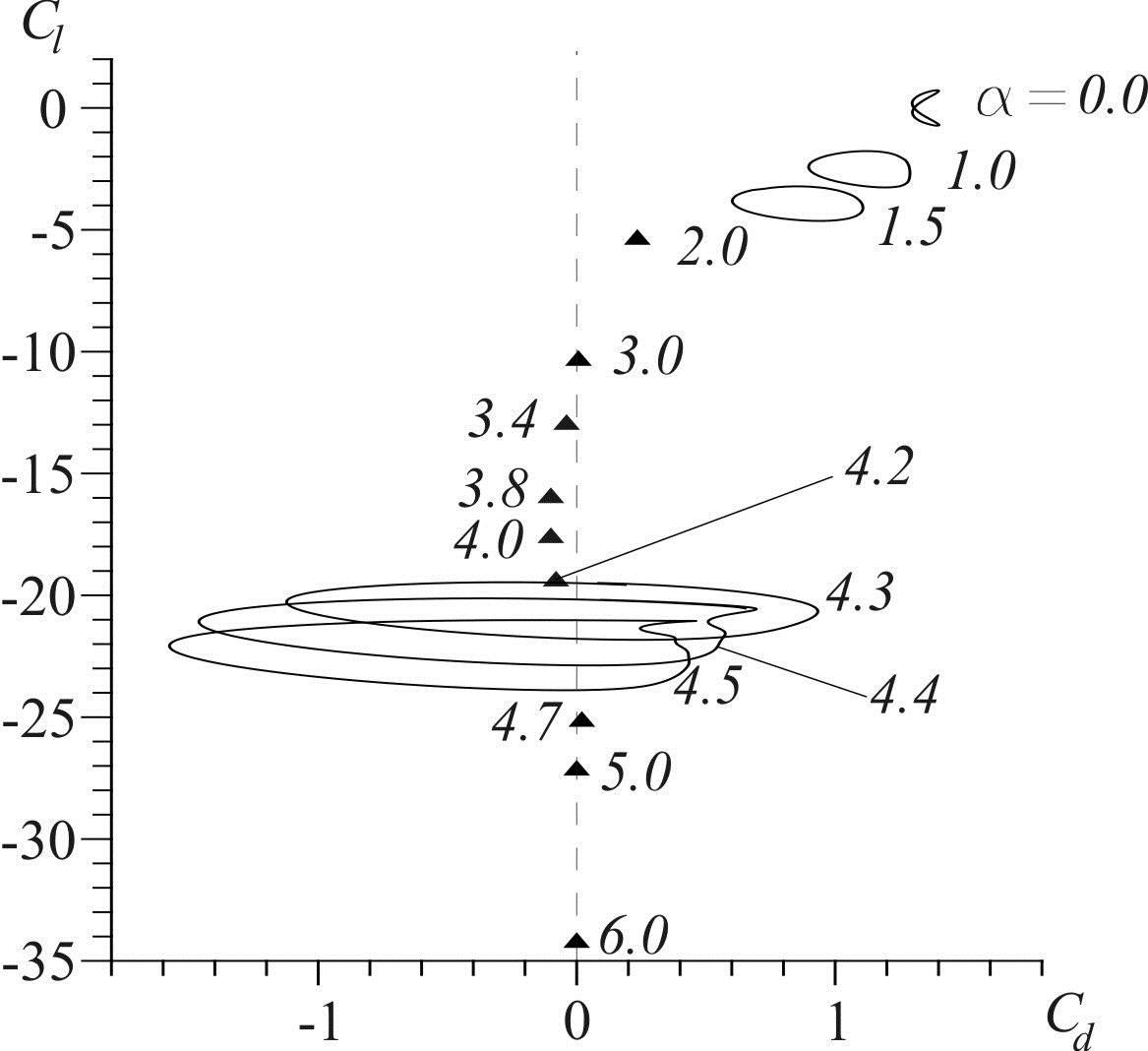}
\caption{Phase diagram $c_d - c_l$ at $\mathsf{Re}=200$.\label{pdiag}}
\end{center}
\end{figure}

\noindent
The changing of flow regimes with the increase in $\alpha$ conveniently can be represented in a phase  diagram (Fig. \ref{pdiag}),where the drag coefficient to the abscissa axis, and the lift coefficients is presented on the ordinate axis. The closed curves correspond to the variations of $c_d$ and $c_l$ in the case of periodic solutions; the points corresponds to their steady values in stationary regimes. It is seen that in the second periodic regime the changes of these coefficients are sufficiently wider in the case of the Karman street at $\alpha<\alpha_L$.

\begin{figure}[h!]
\begin{center}
\includegraphics[width=0.68\textwidth]{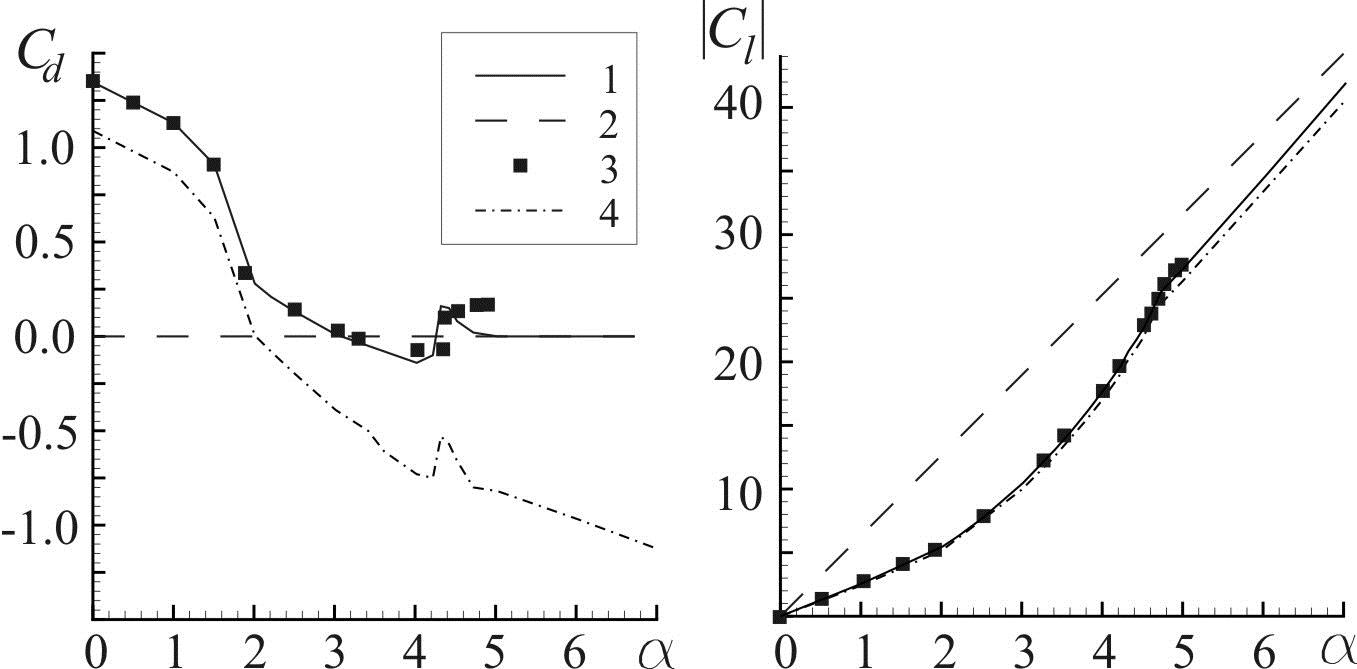}
\caption{Dependence of coefficients $c_d$ and $c_l$ on rotation rate $\alpha$ at $\mathsf{Re}=200$: 1, our calculation; 2, potential flow; 3, calculations (data from Mittal and Kumar, $2003$); 4,component of pressure forces $c^{p}_{d}$ and $c^{p}_{l}$ in values $c_d$ and $c_l$.\label{cdclr}}
\end{center}
\end{figure}

In Fig. \ref{cdclr}, the resultant dependencies of the drag and lift coefficients on the rotation rate at $\mathsf{Re}=200$ are given. With an increase in $\alpha$ the drag coefficient decreases up to realization of the second periodic regime and then achieves a negative values. The dependence $c_l(\alpha)$ is of monotonous character and is so close to the linear dependence as a whole. When determining the $c_l$ coefficient, the role of viscous forces have the same order at $\alpha>\alpha_L$ they are directed in opposite directions.

The numerical results presented here agree well with the results obtained by Mittal and Kumar ($2003$) except for the $c_d$ values at large $\alpha$ (the second steady regime). In this case, $c_d \rightarrow 0$ and $c_l \rightarrow 2\pi\alpha$ in our calculations. This confirms the asymptotic form (Moore, $1957$).

\bigskip
\newpage
\noindent{\bf$\mathbf{3.7}$. The Parametric Flow Chart}

\bigskip

\begin{figure}[h!]
\begin{center}
\includegraphics[width=0.68\textwidth]{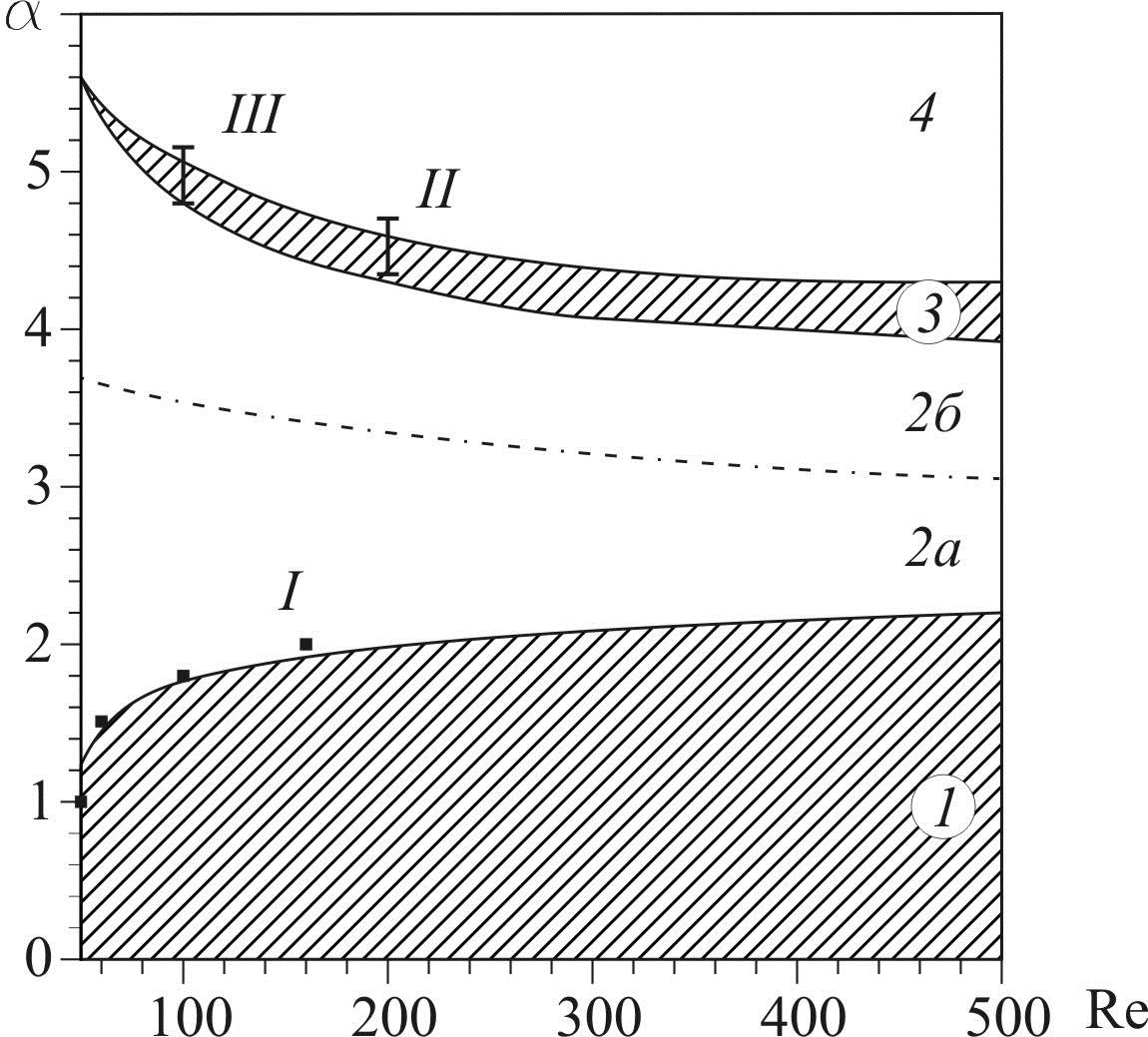}
\caption{Parametric chart of the flow regimes (periodic zones are shaded): $1$, periodic zone with generation of the Karman street; $2a$, steady zone with positive $c_d$; $2b$, steady zone with negative $c_d$; $3$, second periodic zone; $4$, second steady zone.\label{pc}}
\end{center}
\end{figure}

\noindent
The parametric chart of flow regimes is presented in Fig. \ref{pc}. The curve, dividing zone $1$ with the periodic Karman street, agrees with the bifurcation curve (Fig. \ref{pc}, shown by marker I) obtained by Kang et al., ($1999$). The ranges of $\alpha$, in which the second periodic regime is implemented ($4.34\le\alpha\le4.7$ at $\mathsf{Re}=200$) and ($4.8\le\alpha\le5.15$ at $\mathsf{Re}=100$) stated in the works by Mittal and Kumar ($2003$) and Stojkovic et al., ($2002$), correspondigly, are shown in Fig. \ref{pc} (markers II and III) and lay well within zone $3$, which is constructed using the results of the calculations.

The calculations showed that the second periodic zone may be obtained for the same conditions for the Reynolds number as in the first one ($\mathsf{Re}=200$>47), with the conditions for the rotation rate practically not depending on the Reynolds number at $\mathsf{Re}>200$.

Another region is also singled out in Fig. \ref{pc}, in which the drag coefficient has negative values. Negative $c_d$ are found in many alternative numerical investigations of the problem of flow around a rotating cylinder (Kang et al., $1999$; Stojkovic et al., $2002$; Mittal and Kumar, $2003$). Questions regarding the reasons for initiation of such an effect require individual study; however. it is worth nothing that unlike unsteady zones $1$ and $3$, the region with $c_d<0$ exists also at low Reynolds numbers $\mathsf{Re}<47$.

\bigskip

\noindent{\bf$\mathbf{4}$. Conclusions}

\bigskip

\noindent
The results of numerical simulation for the two-dimensional problem of viscous incompressible fluid flow around a rotating cylinder were presented in this work. A finite element algorithm was proposed for the solution of the unsteady Navier–Stokes equations in the stream function–vorticity formulation jointly with the additional non-local boundary condition.

On basis of the calculations performed in the ranges $50\le\mathsf{Re}\le500$ and $0\le\alpha\le7$,a parametric chart of flow regimes was constructed, which demonstrated that the features of flow regime variation slightly depended on the Reynolds number at $\mathsf{Re}>100$. Depending on rotation rate $\alpha$, two steady and two periodic solutions of the problem were revealed.

The asymmetric vortex shedding is generated at negligible $\alpha$ in the wake behind the body. With the achievement by $\alpha$ of critical value $\alpha_L\approx 2$ shedding is suppressed, and the steady solution is realized. Further increase of $\alpha$ leads to an increase in the lift coefficient and a decrease of the drag coefficient; the drag coefficient may even achieve negative values.

In a small range near $\alpha\approx 4.3$ the flow becomes periodic again, with the vortex separation occurring from the upper edge of the cylinder and the drag coefficient increasing.

The further increase of the rotation rate ($\alpha>\alpha_H\approx 4.5$) leads to the steady flow and its pattern resembles the solution of the problem for circulating flow around the cylinder in the frame of potential theory, and the drag coefficient tends to zero and the lift infinitely increases ($c_l=2\pi\alpha$).

\bigskip

\bigskip

\noindent{\bf REFERENCES}

\bigskip

\noindent
{\small Chen, Y.-M., Ou, Y.-R., Pearlstein, A.~J., Development of the wake behind a circular cylinder impulsively started into rotary and rectilinear motion, {\it J. Fluid Mech.,} vol. {\bf 253}, pp. 449-484, 1993.
}

\vspace{0.29cm}
\noindent
{\small Choutanceau, M. and Menard, C., Influence of rotation on the near-wake development behind an impulsively started circular cylinder, {\it J. Fluid Mech.,} vol. {\bf 158}, pp. 399-446, 1985
}

\vspace{0.29cm}
\noindent
{\small Fletcher, C.~A. J.,{\it Computational Techniques for Fluid Dynamics,} vol. {\bf 2}, 2nd ed., New York: Springer, 1991.}

\vspace{0.29cm}
\noindent
{\small Glowinski, R., Finite element methods for incompressible viscous flow, in {\it Handbook of Numerical Analysis,} vol. {\bf 9}, Amsterdam: North-Holland, 2003.
}

\vspace{0.29cm}
\noindent
{\small Kalinin, E.~I. and Mazo A.~B., Numerical simulation of flow around a system of bodies in stream function - vorticity variables, {\it Kazan. Gos. Univ. Uchen. Zap. Ser. Fiz-Mat. Nauki,} vol. {\bf 151}, no. 3, pp. 144-153, 2009 (in Russian).}

\vspace{0.29cm}
\noindent
{\small Kang, S., Choi, H., and Lee, S., Laminar flow past a rotating circular cylinder, {\it Phys. Fluids,} vol. {\bf 11}, no. 11, pp. 3312 - 3321, 1999.}

\vspace{0.29cm}
\noindent
{\small Kuzmin, D. and Turek, S., High-resolution FEM-TVD schemes based on a fully multidimensional flux limiter, {\it J.Comput. Phys.,} vol. {\bf 198}, pp. 131-158, 2004.}

\vspace{0.29cm}
\noindent
{\small Lam, K. M., Vortex shedding flow behind a slowly rotating circular cylinder, {\it J.Fluids Struct.,} vol. {\bf 25}, pp. 245-262, 2009.}

\vspace{0.29cm}
\noindent
{\small Loytsanskiy, L.~G., {\it Mechanics and Gas,} Moscow: State Publisher of Technical and Theoretical Literature, 1950 (in Russian).}

\vspace{0.29cm}
\noindent
{\small Mazo A.~B. and Dautov, R.~Z., On the boundary conditions for the Navier-Stokes equations in stream function - vorticity variables in simulation of a flow around a system of bodies, {\it J. Eng. Phys. Thermophys.,} vol. {\bf 78}, no. 4, pp. 769-776, 2005.}

\vspace{0.29cm}
\noindent
{\small Mazo, A.~B. and Morenko, I.~.V., Numerical simulation of a viscous separation flow around a rotating circular cylinder, {\it J. Eng. Thermophys.,} vol. {\bf 79}, no. 3, pp. 496-502, 2006.}

\vspace{0.29cm}
\noindent
{\small Mittal, S. and Kumar, B., Flow past a rotating cylinder, {\it J. Fluid Mech.,} vol. {\bf 476}, p. 303-334, 2003.
}

\vspace{0.29cm}
\noindent
{\small Mittal, S., Three-dimensional instabilities in flow past a rotating cylinder, {\it J. Fluid. Mech.,} vol. {\bf 71}, pp. 89-95, 2004.
}

\vspace{0.29cm}
\noindent
{\small Moore, D.~W., The flow past a rapidly rotating circular cylinder in a uniform stream, {\it J. Fluid. Mech.,} vol. {\bf 2}, pp. 541-550, 1957.
}

\vspace{0.29cm}
\noindent
{\small Orlanski, I., A simple boundary condition for unbounded hyperbolic flows, {\it J. Comput. Phys.,} vol. {\bf 21}, no. 3, pp. 251-269, 1976.
}

\vspace{0.29cm}
\noindent
{\small Prandtl, L., The Magnus effect and wind-poered ships, {\it Naturwissenschaften,} vol. {\bf 12} pp. 93-108, 1925.
}

\vspace{0.29cm}
\noindent
{\small $Prihod^{,}ko$, A.~A. and Redtchits, D.~A., Numerical modeling of a viscous incompressible unsteady separated flow past a rotating cylinder, {\it Fluid Dyn.,} vol. {\bf 14}, no. 6, pp. 823-829, 2009.
}

\vspace{0.29cm}
\noindent
{\small Stojkovic, D., Breuer, M., and Durst, F., Effect of high rotation rates on the laminar flow around a circular cylinder, {\it Phys. Fluids,} vol. {\bf 14}, no. 9, pp. 3160-3178, 2002.
}

\vspace{0.29cm}
\noindent
{\small Tokumaru, P.~T. and Dimotakis, P.~E., Rotary oscillation control of cylinder wake, {\it J. Fluid Mech.,} vol {\bf 224}, pp. 77-90, 1991.
}

\vspace{0.29cm}
\noindent
{\small Trottenberg, U., Oosterlee, C., and Shuller, A., Multigrid, London: Academic, 2001.
}

\vspace{0.29cm}
\noindent
{\small Zdravkovich, M.~M., {\it Flow around Circular Cylinders,} New York: Oxford University Press, 1997.
}

\end{document}